\documentclass[proceedings]{JHEP3}

\PrHEP{PrHEP hep2001}                   % Do not change this one
\conference{International Europhysics Conference on HEP}                % Do not change this one

\usepackage{epsfig}                   % please use epsfig for figures
\def\etal    {\it et al.\rm}

\title{New CLEO~II Results on Charm Hadron Decays}

\author{\speaker{K.K. Gan}\thanks{Representing the CLEO Collaboration.}\\
        Department of Physics,
        The Ohio State University,
        Columbus, OH 43210,
        U.S.A.\\
        E-mail: \email{gan@mps.ohio-state.edu}}

\abstract
{We present new results on charm hadron decays from the CLEO~II experiment.
The $D^{*+}$ width is measured for the first time, $\Gamma(D^{*+}) = 96 \pm 4 \pm 22$~keV.
The semileptonic decay branching fractions are determined to be
$B(D^+ \to \bar{K}^{*0}e^+\nu_e) = (6.7 \pm 0.4 \pm 0.5 \pm 0.4)\%$ and
$B(D^+ \to \bar{K}^{*0}\mu^+\nu_\mu) = (6.5 \pm 0.9 \pm 0.5 \pm 0.4)\%$.
We observe evidence for $\Omega_c \to \Omega^- e^+\nu_e$ and measure the product
of the branching fraction and cross section, $B(\Omega_c \to \Omega^- e^+\nu_e) \cdot
\sigma(e^+e^- \to \Omega_cX) = 42.2 \pm 14.1 \pm 11.9$~fb.
Within the framework of Heavy Quark Effective Theory, we measure in the decay
$\Lambda^+_c \to \Lambda e^+\nu_e$ the form factor ratio,
$R = f_2/f_1 = -0.31 \pm 0.06 \pm 0.06$.
This provides strong evidence that the form factor $f_2$ describing spin interaction
of the $s$ quark is non-zero.
The $\Xi^+_c$ lifetime is measured to be $503 \pm 47 \pm 18$~fs.
The results on branching fractions, cross section, and form factor ratio are preliminary.}

\begin{document}

We present new results on charm hadron decays from the CLEO~II
experiment~\cite{CLEO} at the Cornell Electron Storage Ring (CESR).
The data were collected at the $\Upsilon(4S)$
($\sqrt{s} \sim 10.6$ GeV) resonance and at energies just below.
In this paper, we begin with the presentation of the first measurement of the $D^{*+}$ width.
We then report the measurement of the branching fractions for the semileptonic decays
$D^+ \to \bar{K}^{*0}e^+\nu_e$ and $\bar{K}^{*0}\mu^+\nu_\mu$.
We then present evidence for the decay $\Omega_c \to \Omega^- e^+\nu_e$ and
measurement of the product of the branching fraction and cross section.
This is followed by the measurement of form factor ratio in the decay
$\Lambda^+_c \to \Lambda e^+\nu_e$ within the framework of Heavy Quark Effective Theory.
Finally, we present the measurement of $\Xi^+_c$ lifetime.

\section{First Measurement of $D^{*+}$ Width}

A measurement of $D^{*+}$ width provides a test of the non-perturbative
strong physics involving heavy quarks.
The predictions for the width range from 15 to 150 keV~\cite{Belyaev}.
The level splittings in the $B$ sector are not large enough for strong transitions.
Therefore a measurement of the $D^{*+}$ width gives unique information
about the strong coupling constant in heavy-light meson systems.
The width depends only on $g$, a universal strong coupling between heavy
vector and pseudoscaler mesons to the pion, since the small contribution
of the electromagnetic decay can be neglected, yielding
\begin{equation}
\Gamma(D^{*+}) = \frac{g^2}{12\pi f^2_\pi}(2 p^3_{\pi^+} + p^3_{\pi^0}),
\end{equation}
where $f_\pi$ is the pion decay constant and the momentum is for the
pion in the $D^{*+}$ rest frame~\cite{Wise} for the decay $D^{*+} \to D^0\pi^+$
and $D^+\pi^0$, respectively.

\EPSFIGURE{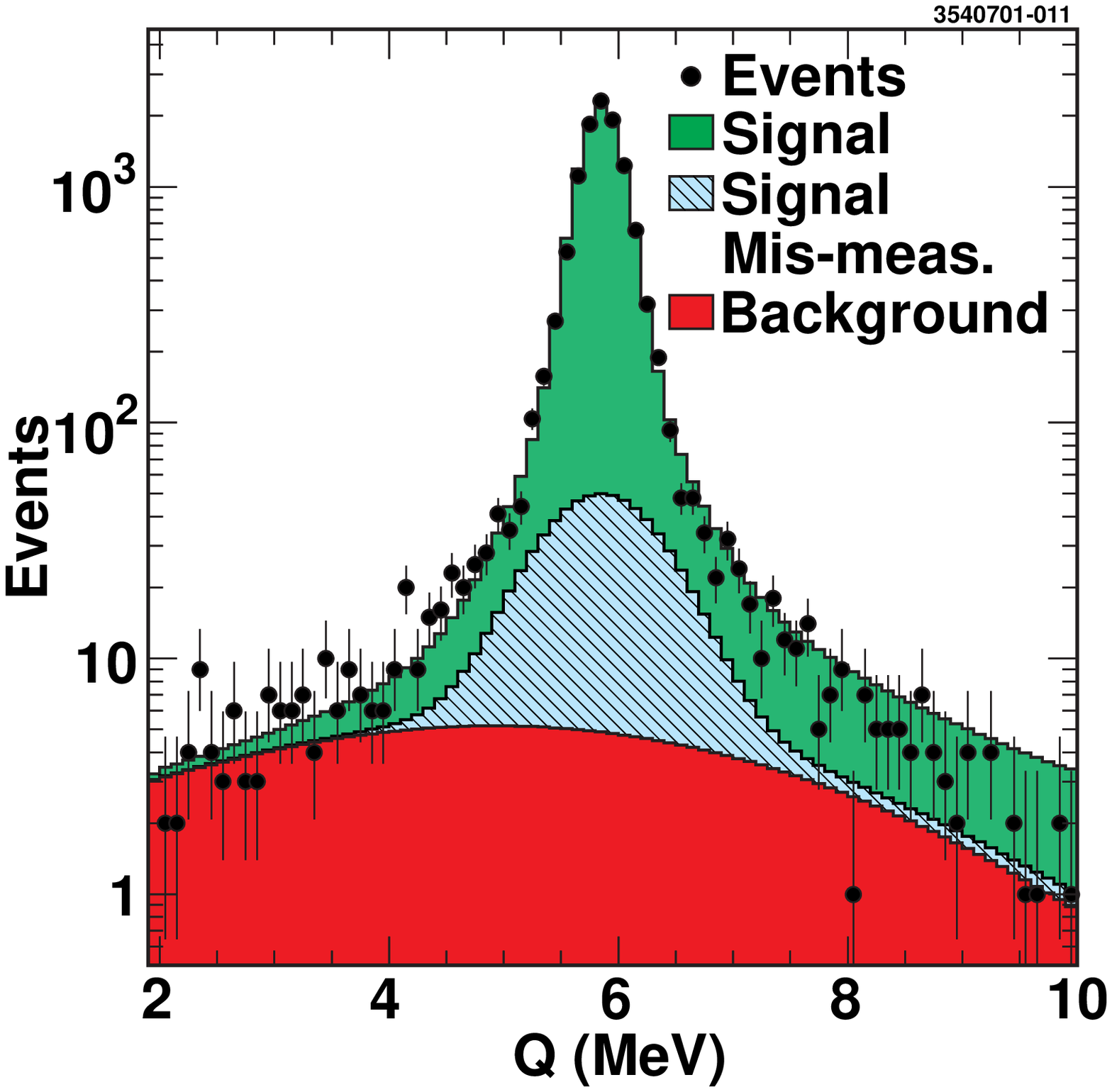,width=4.0cm}                       
{Fit to the energy release distribution in $D^{*+}$ decay.
\label{dstar_width}}

The width is measured from the distribution of the energy release,
$Q \equiv M(K^-\pi^+\pi^+_{slow}) - M(K^-\pi^+) - m_{\pi^+}$, in the decay
$D^{*+} \to \pi^+_{slow}D^0 \to \pi^+_{slow}K^-\pi^+$~\cite{dstar_width_paper}.
The analysis uses a data sample corresponding to an integrated luminosity of
$9.0~\rm fb^{-1}$.
The measurement uses advanced tracking techniques and a reconstruction method
that takes advantage of the small vertical size of the CESR beam spot, a Gaussian
width of $\sim 10~\mu$m vertically as compare to $\sim 300~\mu$m horizontally.
The $K^-$ and $\pi^+$ are required to form a common vertex.
The resultant $D^0$ are then projected back to the CESR luminous region to
determine the $D^0$ production point.
This procedure determines a precise $D^0$ production point for $D^0$'s moving
out of the horizontal plane and hence those moving within 0.3 radians of the
horizontal plane are rejected.
The track $\pi^+_{slow}$ is refit constraining its trajectory to intersect
the $D^0$ production point.
This improves the resolution of the energy release by more than 30\%.
The improvement is essential for this analysis as the resolution of
$\sim$~150~keV is comparable with the $D^{*+}$ width.
The $Q$ distribution as shown in Fig.~\ref{dstar_width} is fit with a P-wave
Breit-Wigner using the unbinned maximum likelihood technique.
The result is
\begin{equation}
\Gamma(D^{*+}) = 96 \pm 4 \pm 22~\rm keV.
\end{equation}
The difference in the result between relativistic and non-relativistic
Breit-Wigner is negligible.
This is the first measurement of the $D^{*+}$ width; the previous upper limit
as measured by the ACCMOR collaboration is $\Gamma(D^{*+}) < 131$~keV
at the 90\% confidence level~\cite{ACCMOR}.
The width yields the strong coupling
\begin{equation}
g = 0.59 \pm 0.01 \pm 0.07.
\end{equation}
This is consistent with theoretical predictions based on HQET and relativistic
quark models, but higher than predictions based on QCD sum rules.
We also measure the energy release in the decay and find
\begin{equation}
\Delta m \equiv m_{D^{*+}} - m_{D^{0}} = 145.412 \pm 0.002 \pm 0.012~{\rm MeV}/c^2.
\end{equation}
This agrees with the value, $145.436 \pm 0.016~{\rm MeV}/c^2$, computed
from a global fit of all flavors of $D^*-D$ mass difference~\cite{PDG}.

\section{Measurement of Branching Fractions for
$D^+ \to {\bar K^{*0}}l^+\nu_l$}

In the semileptonic decay $D^+ \to {\bar K^{*0}}l^+\nu_l$, the transition
amplitude is proportional to the product of the leptonic and hadronic
currents.
The hadronic current can be parameterized by a few analytic functions
called form factors.
There are no reliable calculations of the form factors.
A measurement of the form factors will help to guide the theoretical progress.
It will also help to reduce the uncertainty in the measurement of $V_{ub}$
from the decay $b \to ul\nu$ since the form factors for the two decays
are related.

\EPSFIGURE{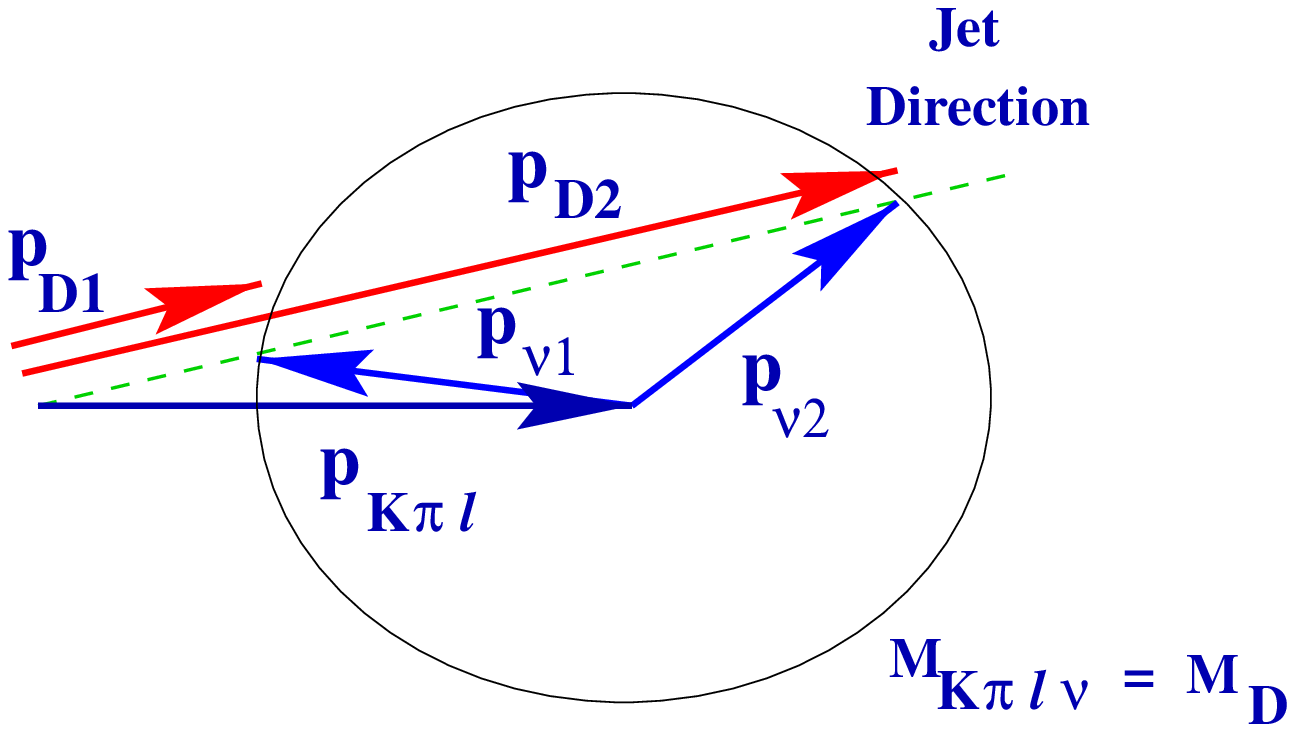,width=4.0cm}
{Locus that provides the solutions for the two $\nu$ momenta in
the decay $D^+ \to {\bar K^{*0}}l^+\nu_l$.\label{locus}}

We measure the semileptonic decay branching fractions using the decay chain
$D^{*+} \to D^+\pi^0$, $D^+ \to {\bar K^{*0}}l^+\nu_l$, and
${\bar K^{*0}} \to K^-\pi^+$.
The data sample used corresponds to an integrated luminosity of 9.0~fb$^{-1}$.
The analysis assumes that the direction of the $D$ meson can be
approximated by the thrust axis of the event~\cite{thrust}.
Two methods are used to obtain up to three values for the missing
$\nu$ momentum, $\vec{p}_\nu$.
In the first method, the two $\nu$ momenta are given by the two intersections
of the $D$ direction with the locus satisfying $m_{K\pi l\nu} = m_{D^+}$
as shown in Fig.~\ref{locus}.
The second method uses the missing momentum of each event as an estimate
of the $\nu$ momentum.
Among these three $\nu$ momentum estimates, the estimate which gives
$\delta m = m_{K\pi l\nu\pi^0} - m_{K\pi l\nu}$ closest to the
known value of $m_{D^{*+}} - m_{D^+}$~\cite{PDG} is chosen.
To measure the number of candidate semileptonic decays, we first plot
$K\pi$ mass distribution in bins of $\delta m$ and then fit for the
number of $K^{*0}$.
We then plot the $K^{*0}$ yield as a function of $\delta m$ and fit
the resultant distribution for the number of semileptonic decay candidates.
Figure~\ref{deltam} shows the $\delta m$ distribution of the
electron mode.

\EPSFIGURE{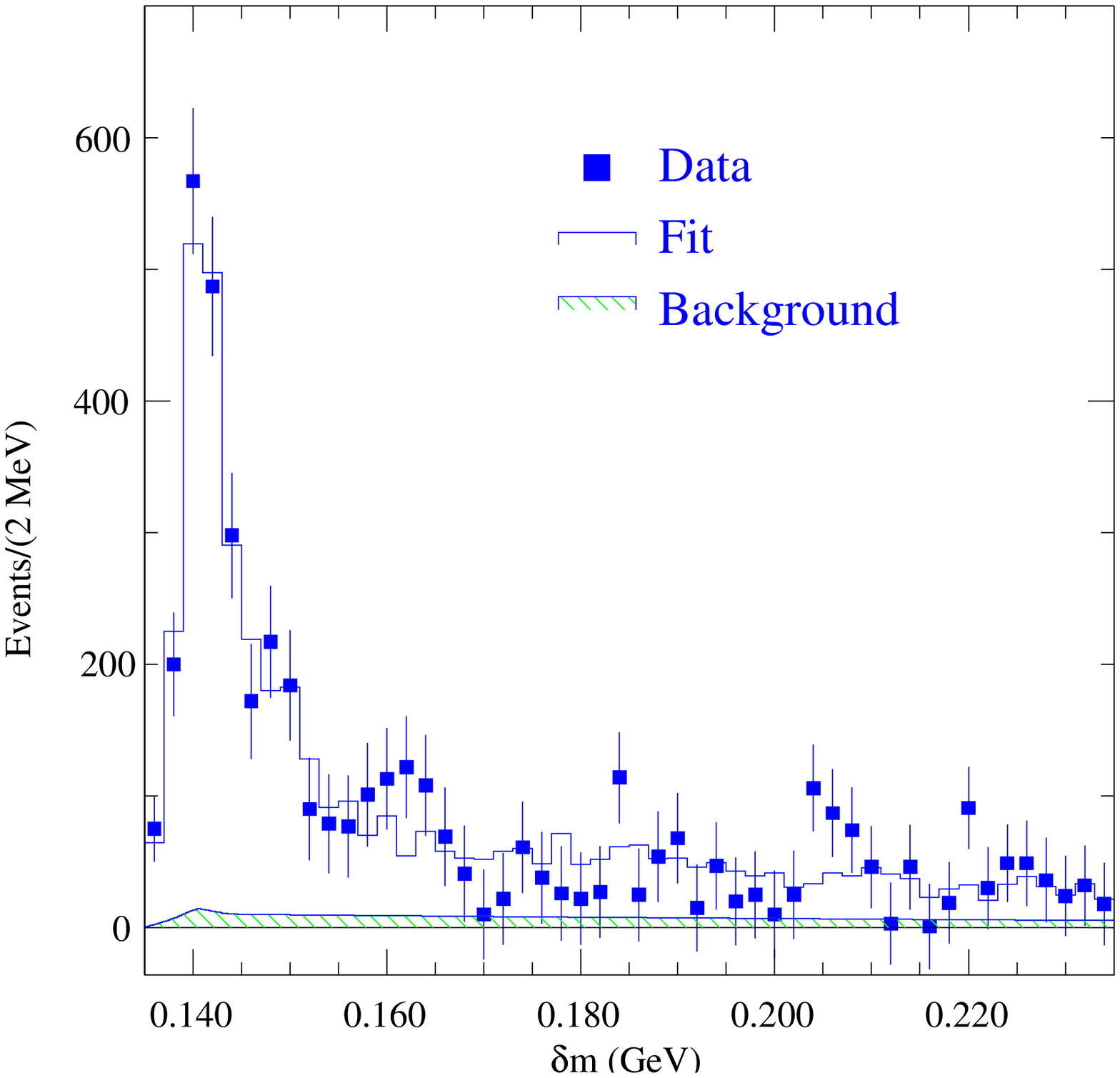,width=4.0cm}
{$\delta m$ distribution of the decay $D^+ \to {\bar K^{*0}}e^+\nu_e$.
\label{deltam}}

The semileptonic branching fractions are measured by normalizing
to the decay $D^+ \to K^-\pi^+\pi^+$ for two reasons.
First, the branching fraction of the normalization decay has been measured
with good precision and second, many systematic errors cancel in the ratio.
The normalization decay candidates are selected with similar selection
criteria and the yield extracted using a similar $\delta m$ fitting procedure.
The results on the branching fractions are
\begin{eqnarray}
B(D^+ \to \bar{K}^{*0}e^+\nu_e) &=& (6.7 \pm 0.4 \pm 0.5 \pm 0.4)\%\\
B(D^+ \to \bar{K}^{*0}\mu^+\nu_\mu) &=& (6.5 \pm 0.9 \pm 0.5 \pm 0.4)\%.
\end{eqnarray}

This is the first simultaneous measurements of both branching fractions
by the same experiment.
The two measurements are consistent with lepton universality.
The measurements are also consistent with theoretical predictions~\cite{Isgur}
and other measurements~\cite{PDG} except for the electron mode which is
$\sim 3\sigma$ higher than that of E691~\cite{E691}.

\section{Evidence for the Decay $\Omega_c \to \Omega^- e^+\nu_e$}

\EPSFIGURE{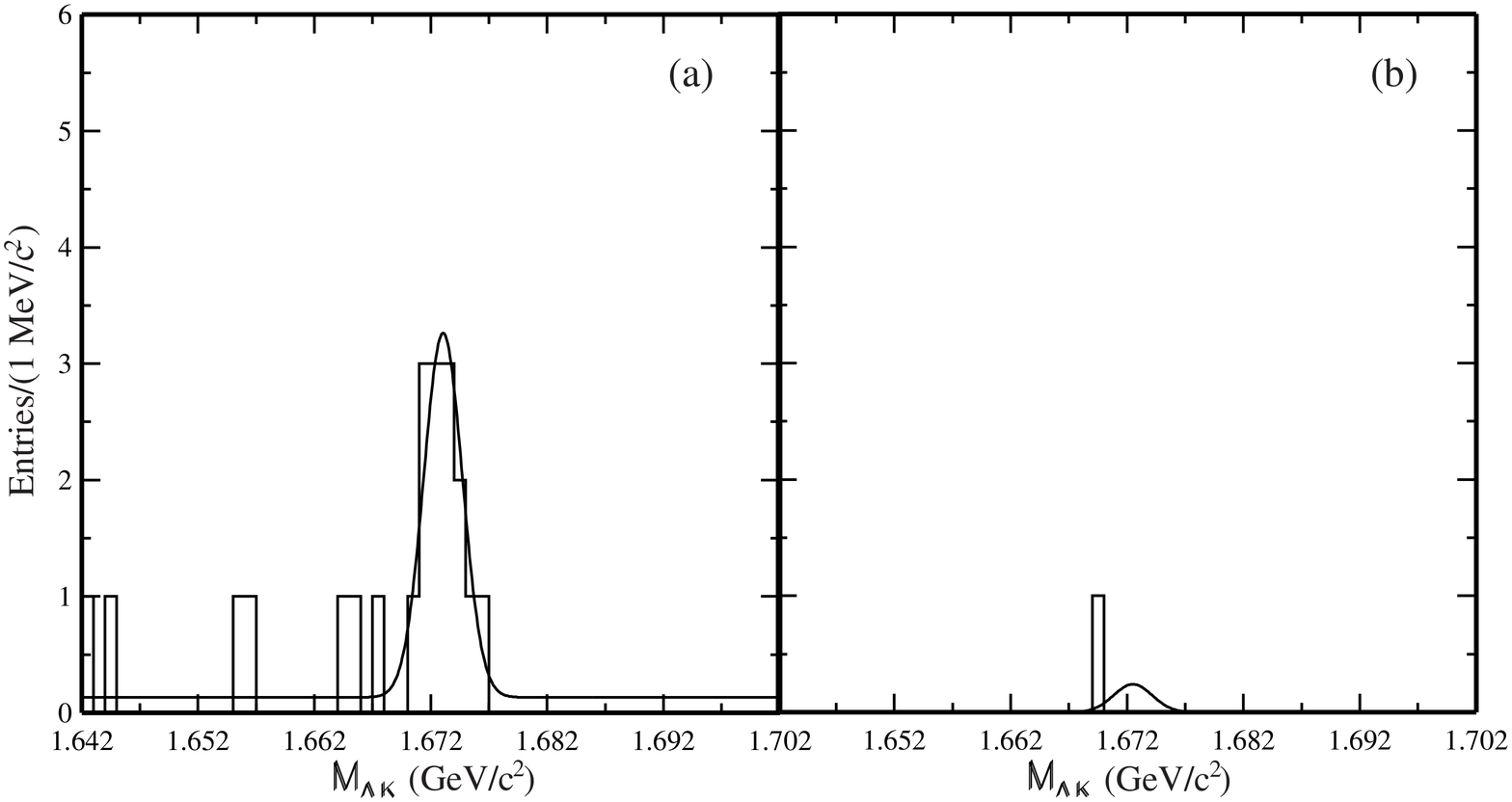,width=6.70cm}                       
{Invariant mass of $\Lambda K$ pairs with a right (a) and wrong (b) sign lepton.
\label{omega}}

Heavy Quark Expansions (HQE) provide a successful description of the 
lifetimes of charm hadrons and of the absolute semileptonic branching ratios
of the $D^0$ and $D_s$~\cite{Bigi}.
Charm baryons provide new information as corrections to HQE of order $1/m_c^2$
and $1/m_c^3$, yielding sizeable differences between the semileptonic widths of
charm mesons and  baryons.  
Comparing the semileptonic widths therefore provides a test of HQE and duality. 

\EPSFIGURE{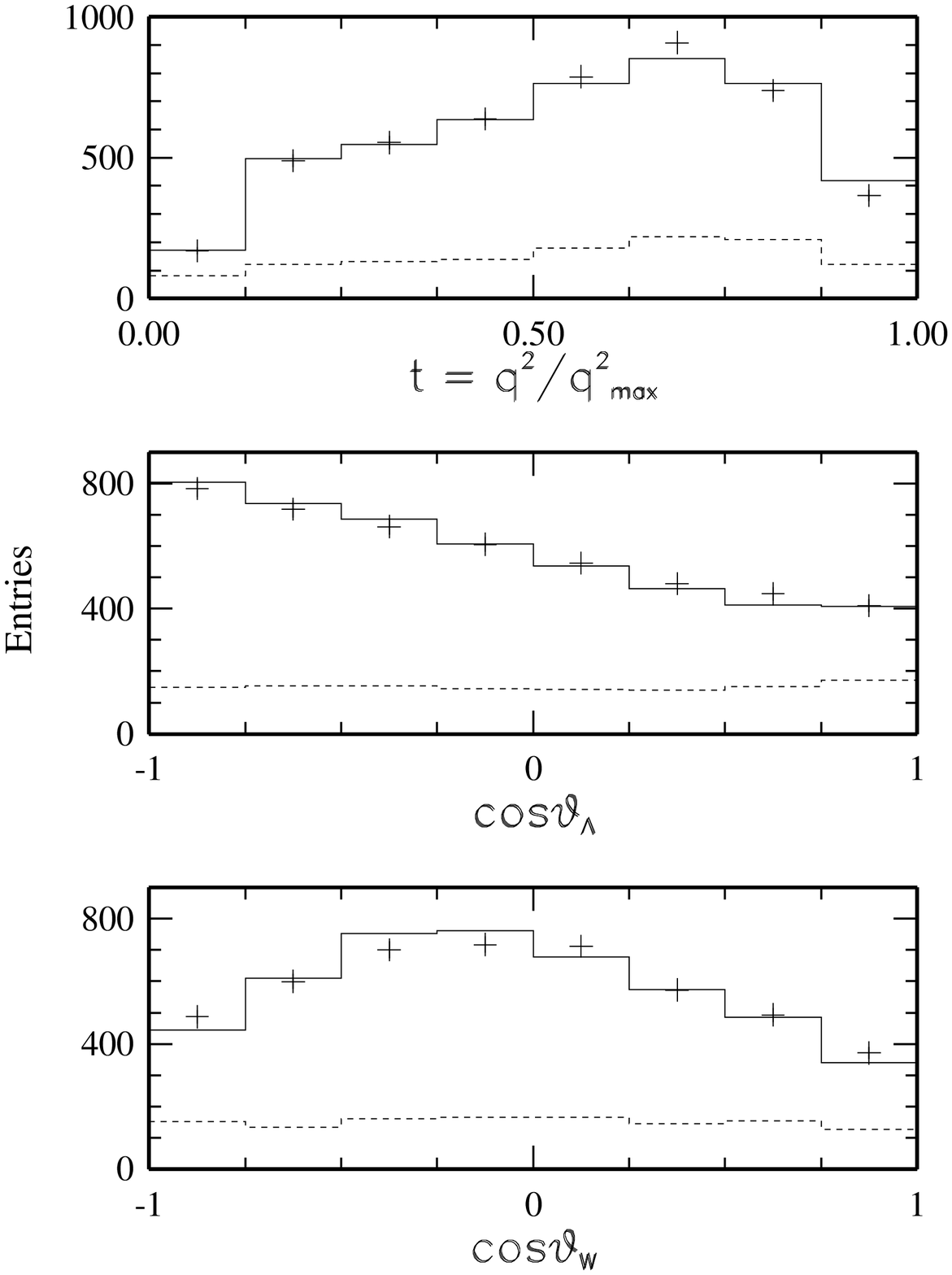,width=6.2cm}                       
{Projections of the fit (solid lines) to the kinematic variables of
the decay $\Lambda^+_c \to \Lambda e^+\nu_e$.
The dashed lines are the sum of the backgrounds.
\label{lambda}}

We searched in $e^+ e^- \rightarrow c \bar{c}$ events for the decay
$\Omega_c \rightarrow \Omega^- e^+ \nu_e$ with
$\Omega^- \rightarrow \Lambda K^{-}$ and $\Lambda \rightarrow p\pi^{-}$.
We identify the decay by detecting $\Omega^- e^+$  combinations
with invariant mass in the range $ m_{\Omega} < m_{\Omega e} < m_{\Omega_c}$. 
The analysis is based on a data sample with an integrated luminosity of 13.8~fb$^{-1}$.
Figure~\ref{omega} shows the $\Lambda K$ mass distribution for combinations with
a right sign ($e^+$) and wrong sign ($e^-$) lepton.
A fit to the mass spectrum yields $13.0 \pm 3.8$
$\Omega^- e^+$ pairs of which $11.4 \pm 3.8$ are consistent with the decay
$\Omega_c \rightarrow \Omega^- e^+ \nu_e$ and the remainder are consistent with background.
Only 1 wrong sign pair is observed.
The probability that the signal is due to a 
background fluctuation is $P < 9 \times 10^{-4}$.
We have therefore established evidence for the semileptonic decay
$\Omega_c \rightarrow \Omega^- e^+ \nu_e$.
This is the first $\beta$ decay where none of the quarks in the parent baryon are light.
The product of the branching fraction and cross section is measured to be
\begin{equation}
B(\Omega_c \to \Omega^- e^+\nu_e) \cdot
\sigma(e^+e^- \to \Omega_cX) = 42.2 \pm 14.1 \pm 11.9\rm~fb.
\end{equation}

\section{Form Factor Ratio Measurement in the Decay $\Lambda^+_c \to \Lambda e^+\nu_e$}

In HQET if the initial and final states of $\Lambda$ type baryons both contain
a heavy quark, only one form factor governs the semileptonic transition.
If the quark produced in the decay is light, there are two form factors.
The second form factor is expected to be smaller and it accounts for the spin
interaction between the quarks in the $\Lambda$.
By studying the decay rate distribution of the decay kinematic variables
in the decay $\Lambda^+_c \to \Lambda e^+\nu_e$, we extract the ratio
$R = \frac{f_2}{f_1}$ assuming a dipole dependence of the form factor evolution~\cite{Korner}.
The same set of form factors governs the decay $\Lambda_b \rightarrow p e^- \bar{\nu_e}$
and hence our measurement provides an important input for the extraction of $|V_{ub}|$
from the decay.

We searched for $\Lambda_c$ in $e^+ e^- \rightarrow c \;\overline{c}$ events
by detecting $\Lambda e^+$ pairs with $\Lambda \to p\pi^-$.
The analysis uses a data sample with an integrated luminosity of 13.4~fb$^{-1}$.
The momentum of $\Lambda_c$ cannot be measured due to the undetected neutrino.
An estimate of the momentum is obtained by requiring
consistency of the $\Lambda$ and $e^+$ with originating in a semileptonic
decay, the overall event shape, and the $\Lambda_c$ fragmentation function.
The form factor ratio is measured from the decay rate distribution
of three decay kinematic variables, $t = q^2/q^2_{max}$, $\cos\theta_\Lambda$, and
$\cos\theta_W$, where $\theta_\Lambda$ ($\theta_W$) is the angle between
$p$ and $\Lambda$ ($e$ and $W$) in the center of mass of the $\Lambda$ ($W$).
An unbinned maximum likelihood fit~\cite{UCSB} to the kinematic variable distributions
(see Fig.~\ref{lambda}) yields
\begin{equation}
R = f_2/f_1 = -0.31 \pm 0.06 \pm 0.06.
\end{equation}
This is consistent with our previous measurement~\cite{Crawford} but with
much improved precision.
The result provides strong evidence that the form factor $f_2$ describing the
spin interaction of the $s$ quark is non-zero.

\section{Measurement of $\Xi^+_c$ Lifetime}

Charm baryon lifetime measurements provide insight into the dynamics of
non-perturbative heavy quark decays.
Unlike the case of charm mesons, the exchange mechanism is not helicity suppressed
and therefore can be comparable in magnitude to the spectator diagram.
In addition, color suppression is only active for particular decay channels.
While several theoretical models~\cite{Shifman} can account for the apparent lifetime
hierarchy, $\tau_{\Xi^+_c} > \tau_{\Lambda^+_c} > \tau_{\Xi^0_c} > \tau_{\Omega_c}$,
experimental results are necessary to advance our understanding of the various
contributions to the hadronic width.

\EPSFIGURE{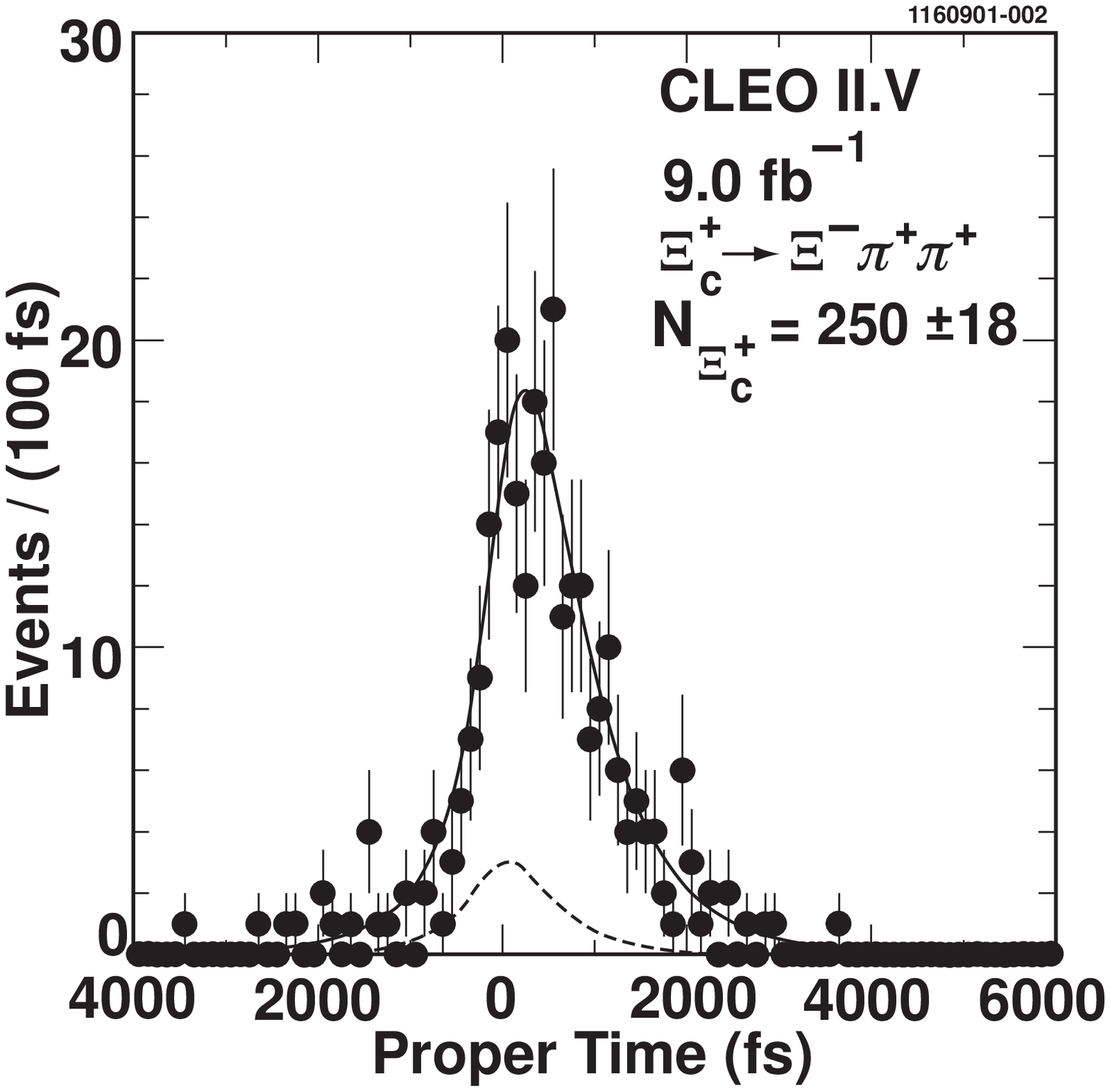,width=4.0cm}                       
{Proper time distribution of $\Xi^+_c$ candidates.
\label{lifetime}}

We measure the $\Xi^+_c$ lifetime using the decay $\Xi^+_c \to \Xi^-\pi^+\pi^+$
with $\Xi^- \to \Lambda\pi^-$ and $\Lambda \to p\pi^-$~\cite{Xi_c_paper}.
The lifetime is measured from the decay length in the vertical direction because
of the significantly smaller beam spread (see Section 1).
The measurement takes advantage of the three-layer double-sided silicon strip
tracker installed in the CLEO~II.V detector.
The analysis uses a data sample corresponding to an integrated luminosity of
$9.0~\rm fb^{-1}$.
The proper time distribution of the $\Xi^+_c$ candidates is shown in
Fig.~\ref{lifetime}.
An unbinned maximum likelihood fit yields,
\begin{equation}
\tau_{\Xi^+_c} = 503 \pm 47 \pm 18~\rm fs.
\end{equation}
This is the first measurement of the lifetime from an $e^+e^-$ experiment
with different systematic errors from those of the fixed target experiments.
The result is higher than the current world average, $330^{+60}_{-40}$~fs, as
is the Focus collaboration's result of $439 \pm 22 \pm 9$~fs~\cite{Focus}.
We can combine our result with the recent CLEO~II.V measurement~\cite{Lambda_c_paper}
of the $\Lambda^+_c$ lifetime, $\tau_{\Lambda^+_c} = 179.6 \pm 6.9 \pm 4.4$~fs
to obtain $\tau_{\Xi^+_c}/\tau_{\Lambda^+_c} = 2.8 \pm 0.3$.
The CLEO~II.V ratio is higher than the expectations, $\sim 1.2-1.7$, from
models based on a $1/m_c$ expansion~\cite{Shifman}.

\acknowledgments
This work was supported in part by the U.S.~Department of Energy.
The author wishes to thank M. Dubrovin, T. Hart, S.J. Lee, and V. Pavlunin
for helps in preparing the talk.
The author wishes to thank R.~Kass for the careful reading of this manuscript.

% \TABLE{\begin{tabular}...
%       ....
%       \end{tabular}%
%       \caption{Text of the caption             % If you need to put
%                of the table.\label{tablabel}}} % a table

\end{document}